\documentclass[aps,prd,onecolumn,notitlepage,nofootinbib,superscriptaddress,11pt,longbibliography]{revtex4-1}

\usepackage[utf8]{inputenc}
\usepackage[T1]{fontenc}
\usepackage{amsmath,amssymb,amsfonts}
\usepackage{graphicx}
\usepackage{hyperref}
\usepackage{xcolor}
\usepackage{bm}
\usepackage{physics_local}

\hypersetup{
    colorlinks=true, 
    linkcolor=blue, 
    citecolor=blue, 
    urlcolor=blue
}

\makeatletter

\renewcommand*{\@fnsymbol}[1]{\ensuremath{\ifcase#1\or \dagger\or *\or \ddagger\or \mathsection\or \mathparagraph\or \|\or **\or \dagger\dagger \or \ddagger\ddagger \else\@ctrerr\fi}}
\makeatother

\begin{document}

\title{Absence of CP Violation in the Scalar Sector of a Higgs Triplet Model}

\author{Matheo Escobar}
\email{m.escobarvergara@uandresbello.edu} 
\affiliation{Universidad Andres Bello, Departamento de Física y Astronomía, Facultad de Ciencias Exactas, Sazi\'e 2212, Piso 7, Santiago, Chile}

\author{Sebasti\'an Olivares}
\email{sebastian.olivares@uss.cl} 
\affiliation{Facultad de Ingeniería, Universidad San Sebastián, Bellavista 7, Santiago 8420524, Chile}

\date{\today}

\begin{abstract}
We investigate the potential for spontaneous CP violation in a Higgs Triplet Model extension, including a complex singlet scalar. We demonstrate that the scalar potential strictly forbids spontaneous CP violation across the entire parameter space. For a non-zero trilinear coupling $\kappa$, minimization conditions combined with global symmetries enforce a phase alignment that yields a real vacuum. In the singular limit $\kappa \to 0$, we show that the enhancement of global symmetries renders unphysical vacuum phases. Consequently, we conclude that the scalar sector of this model cannot source CP violation.
\end{abstract}

\maketitle
\section{Introduction}

Extensions of the Standard Model (SM) scalar sector can provide natural frameworks to address two of the most pressing problems in particle physics: the origin of neutrino masses and the CP violation required for baryogenesis \cite{Fukuda:1998mi, Sakharov:1967dj, Kobayashi:1973fv, Jarlskog:1985ht}. The Higgs Triplet Model (HTM) that includes the Type-II Seesaw mechanism, is a minimal extension that successfully generates neutrino masses through the introduction of a scalar triplet with hypercharge $Y=1$ \cite{Schechter:1980gr, Cheng:1980qt, Magg:1980ut, Lazarides:1980nt}. This mechanism provides a complementary approach to the canonical Type-I seesaw, allowing for potentially observable signatures at colliders while naturally accommodating small neutrino masses \cite{Arhrib:2011uy, Maki:1962mu}.

However, the vacuum structure of the minimal HTM is often constrained by electroweak precision measurements, particularly the $\rho$ parameter \cite{ParticleDataGroup:2024cfk}, which restricts the triplet vacuum expectation value (VEV) to be small, $v_3 \lesssim \mathcal{O}(1\,\text{GeV})$ \cite{FileviezPerez:2008jbu}. To address issues such as vacuum stability, spontaneous lepton number breaking, or enhanced phenomenological flexibility, the model is frequently extended with a complex singlet scalar, forming the so called ``123 model'' (containing representations of dimension 1, 2, and 3). This setup was first proposed as a framework for spontaneous lepton-number violation, giving rise to a Majoron \cite{Chikashige:1981ui, Gelmini:1981ct, Schechter:1982dc}. Subsequent works have explored its collider phenomenology as well as the properties of its scalar potential and vacuum \cite{FermiophobicHiggs, MAD1998htm}. This extension preserves the Type-II seesaw mechanism while providing additional sources of spontaneous symmetry breaking and a richer scalar spectrum.

A key question in such multi-scalar extensions is the origin of CP violation required for baryogenesis. While spontaneous CP violation (SCPV) through complex vacuum expectation values is physically attractive \cite{Lee:1973iz, Branco:1980sz} and well-established in the Two-Higgs-Doublet Model \cite{Branco:2011iw, Grzadkowski:2016scpv}, triplet-extended models could exhibit more rigid vacuum structures \cite{Ferreira:2021szw}. Does the 123 model's scalar sector provide this CP violation spontaneously?

In this work, we analyze the vacuum structure of the 123 model to determine whether the interplay between the singlet and triplet sectors can source SCPV. Our analysis demonstrates that when the trilinear coupling $\kappa$ connecting the singlet and triplet sectors is non-zero, spontaneous CP violation in the scalar sector is strictly forbidden. As for $\kappa = 0$, the vacuum structure imposes no restriction on the VEV phases, but the enhanced global symmetry allows both VEVs to be rotated to real values. This result arises from the minimization conditions which, combined with the hermiticity of the potential, lock the relative phases of the VEVs to zero (or $\pi$).

This result means that CP violation in the 123 model—often required for leptogenesis \cite{Ma:1998, Buchmuller:2005eh}—cannot arise spontaneously from the vacuum and must instead be introduced explicitly, e.g. through Yukawa interactions and/or complex scalar couplings.

The paper is organized as follows: Section \ref{sec:model} introduces the model and the scalar potential. Section \ref{sec:phase_constraints} derives the phase constraints from minimization. Section \ref{sec:absence_scpv} establishes the absence of SCPV and analyzes the mass spectrum. Section \ref{sec:discussion} discusses the comparison with the 2HDM and implications for leptogenesis. We conclude in Section \ref{sec:conclusions}.

\section{Model and Scalar Potential}
\label{sec:model}

The scalar sector consists of three $SU(2)_L \times U(1)_Y$ multiplets: a complex singlet $\sigma$ ($Y=0$, $L=2$), a doublet $\Phi$ ($Y=-1/2$, $L=0$), and a triplet $\Delta$ ($Y=1$, $L=-2$). The field representations are defined as:
\begin{equation}
\sigma = \sigma, \qquad 
\Phi = \begin{pmatrix} \phi^0 \\ \phi^- \end{pmatrix}, \qquad 
\Delta = \begin{pmatrix} \Delta^0 & \Delta^+/\sqrt{2} \\ \Delta^+/\sqrt{2} & \Delta^{++} \end{pmatrix}.
\end{equation}

The most general renormalizable scalar potential invariant under the gauge and global symmetries is given by:
\begin{equation}
\begin{aligned}
V &= \mu_1^2 \sigma^\dagger \sigma + \mu_2^2 \Phi^\dagger \Phi + \mu_3^2 \text{Tr}(\Delta^\dagger \Delta) \\
&\quad + \lambda_1 (\Phi^\dagger \Phi)^2 + \lambda_2 (\text{Tr}(\Delta^\dagger \Delta))^2 + \lambda_3 \Phi^\dagger \Phi \text{Tr}(\Delta^\dagger \Delta) \\
&\quad + \lambda_4 \text{Tr}(\Delta^\dagger \Delta \Delta^\dagger \Delta) + \lambda_5 (\Phi^\dagger \Delta^\dagger \Delta \Phi) \\
&\quad + \beta_1 (\sigma^\dagger \sigma)^2 + \beta_2 (\Phi^\dagger \Phi)(\sigma^\dagger \sigma) + \beta_3 \text{Tr}(\Delta^\dagger \Delta) \sigma^\dagger \sigma \\
&\quad - \kappa\left(  \Phi^T \Delta \Phi \sigma + \text{h.c.} \right),
\end{aligned}
\label{eq:scalar_potential}
\end{equation}
where the last term represents the trilinear mixing $\kappa$ which acts as a portal between the singlet and triplet sectors, explicitly breaking the individual global symmetries of the fields while preserving the total lepton number $L$.

A detailed analysis of vacuum stability and the mass spectrum is provided in Sec.~3 of Ref.~\cite{MAD1998htm}. In the standard treatment, the neutral components acquire real VEVs $v_1, v_2, v_3$. Defining the CP transformation as $\varphi(x) \xrightarrow{CP} \varphi(-x)^*$, the fields are expanded around the vacuum in terms of CP-even ($R_i$) and CP-odd ($I_i$) eigenstates:
\begin{equation}
\sigma = \frac{v_1 + R_1 + iI_1}{\sqrt{2}}, \quad
\phi^0 = \frac{v_2 + R_2 + iI_2}{\sqrt{2}}, \quad
\Delta^0 = \frac{v_3 + R_3 + iI_3}{\sqrt{2}}.
\end{equation}
With this definition, the $R_i$ fields have CP parity $+1$ (scalars) and the $I_i$ fields have CP parity $-1$ (pseudoscalars). The phenomenology of this real-vacuum scenario, including the scalar mass spectrum and the Type-II seesaw mechanism, has been extensively studied in the literature \cite{FermiophobicHiggs, MAD1998htm}.

\subsection{Generalization to Complex Vacuum}
To study the possibility of spontaneous CP violation, we allow the scalar VEVs to carry complex phases. We therefore parametrize the vacuum as:
\begin{equation}
v_1 \to v_1 e^{i\theta_1}, \qquad v_3 \to v_3 e^{i\theta_3}.
\end{equation}
Note that we have used the $U(1)_Y$ gauge freedom to rotate the doublet VEV to be real ($v_2 \in \mathbb{R}$), transferring any physical CP-violating phases to the singlet and triplet sectors.

\section{Phase Constraints from Minimization}
\label{sec:phase_constraints}

In this section, we derive the constraints imposed by the minimization of the scalar potential on the vacuum phases. The minimization conditions are obtained by requiring the first derivatives of the potential to vanish at the vacuum, $\partial V/\partial \phi_i|_{\text{vac}} = 0$. We parametrize the VEVs in terms of real magnitudes $v_i$ and phases $\theta_i$, while the trilinear coupling $\kappa$ is taken to be real to preserve explicit CP symmetry.

\subsection{Vacuum Parametrization}
The vacuum expectation values correspond to the neutral components of the fields:

\begin{equation}
\langle\sigma\rangle = \frac{v_1}{\sqrt{2}}e^{i\theta_1}, \quad 
\langle\Phi\rangle = \frac{1}{\sqrt{2}}\begin{pmatrix} v_2 \\ 0 \end{pmatrix}, \quad 
\langle\Delta\rangle = \frac{1}{\sqrt{2}}\begin{pmatrix} v_3 e^{i\theta_3} & 0 \\ 0 & 0 \end{pmatrix}.
\end{equation}

The key constraints on the vacuum phases arise from the minimization conditions with respect to the pseudoscalar components $I_1, I_2, I_3$. For a potential where CP is explicitly conserved ($\kappa \in \mathbb{R}$), these yield:

\begin{align}
\frac{\partial V}{\partial I_1} &\propto \kappa {v_2^2 v_3} \sin(\theta_1 + \theta_3) = 0, \label{eq:tad1} \\
\frac{\partial V}{\partial I_2} &\propto \kappa v_1 v_2 v_3 \sin(\theta_1 + \theta_3) = 0, \label{eq:tad2} \\
\frac{\partial V}{\partial I_3} &\propto\kappa {v_2^2 v_1} \sin(\theta_1 + \theta_3) = 0. \label{eq:tad3}
\end{align}

Assuming a non-trivial vacuum where all VEV magnitudes are non-zero ($v_i \neq 0$) and that the sectors are coupled ($\kappa \neq 0$), all three equations reduce to the single redundant condition:

\begin{equation}
\boxed{\theta_1 + \theta_3 = n\pi, \quad n \in \mathbb{Z}.}
\label{eq:final_constraint}
\end{equation}

This condition is central to the phenomenology of the model. It implies that the phases of the singlet and triplet VEVs are not independent free parameters but are strictly correlated by Eq.~(\ref{eq:final_constraint}), reducing the freedom of the vacuum structure.

For completeness, we use the tadpole conditions for the real components, $\partial V/\partial R_i = 0$, to express the quadratic mass terms $\mu_i^2$ as functions of the VEVs and the couplings:

\begin{align}
\mu_1^2 &= -\beta_1 v_1^2 - \tfrac{1}{2}\beta_2 v_2^2 - \tfrac{1}{2}\beta_3 v_3^2 + \frac{1}{2}\kappa \frac{v_2^2 v_3 \cos(\theta_1 +\theta_3)}{v_1 }, \\
\mu_2^2 &= -\lambda_1 v_2^2 - \tfrac{1}{2}\beta_2 v_1^2 - \tfrac{1}{2}(\lambda_3 + \lambda_5) v_3^2 + \kappa v_1 v_3 \cos(\theta_1+\theta_3), \\
\mu_3^2 &= -(\lambda_2 + \lambda_4) v_3^2 - \tfrac{1}{2}\beta_3 v_1^2 - \tfrac{1}{2}(\lambda_3 + \lambda_5) v_2^2 + \frac{1}{2}\kappa \frac{v_2^2 v_1 \cos(\theta_1 + \theta_3)}{v_3} .
\end{align}

These equations determine the quadratic mass parameters $\mu_i^2$ as functions of the vacuum configuration and play no further role in the analysis of CP violation.

\section{Absence of Spontaneous CP Violation in the Scalar Sector}
\label{sec:absence_scpv}

In this section, we establish our central result. The scalar sector of the 123 model cannot exhibit spontaneous CP violation in any region of parameter space. We analyze separately the generic coupled case ($\kappa \neq 0$) and the singular decoupling limit ($\kappa \to 0$), showing that in both cases all vacuum phases can be removed by global symmetry transformations.

\subsection{The Coupled Case: $\kappa \neq 0$}
Consider the coupled case with a non-zero, real trilinear coupling $\kappa$. As derived in Sec.~\ref{sec:phase_constraints}, the minimization of the potential imposes the phase alignment condition in Eq.~\ref{eq:final_constraint}. Although $\theta_1$ and $\theta_3$ may take nonzero values, they are not physical if they can be removed by a symmetry transformation. The Lagrangian possesses a global $U(1)_L$ symmetry corresponding to lepton number conservation, under which the fields transform as:
\begin{equation}
\sigma \to e^{2i\alpha}\sigma, \quad \Delta \to e^{-2i\alpha}\Delta, \quad \Phi \to \Phi.
\end{equation}
We can utilize this freedom to perform a field redefinition with rotation parameter $\alpha = -\theta_1/2$. Under this transformation, the new vacuum phases $\theta'_i$ become:
\begin{align}
\theta'_1 &= \theta_1 + 2\alpha = 0, \\
\theta'_3 &= \theta_3 - 2\alpha = \theta_3 + \theta_1.
\end{align}
Applying Eq.~\eqref{eq:final_constraint}, the new triplet phase becomes $\theta'_3 = n\pi$. Consequently, the vacuum expectation values in the new basis are:
\begin{equation}
\langle\sigma\rangle' \in \mathbb{R}, \qquad \langle\Delta^0\rangle' \propto e^{in\pi} = \pm 1 \in \mathbb{R}.
\end{equation}

Working in a CP-conserving basis where all parameters of the scalar potential are real, we have shown that for $\kappa \neq 0$ the minimization conditions (together with the available rephasings) allow all VEVs to be chosen real. The vacuum therefore preserves CP, and spontaneous CP violation cannot occur in the 123 model when $\kappa \neq 0$.

\subsection{The Decoupling Limit and Physical Consequences}
In the coupled case $\kappa \neq 0$, the phase alignment $\theta_1 + \theta_3 = n\pi$ enforces a block-diagonal neutral scalar mass matrix, preventing CP-even/CP-odd mixing. The CP-odd sector comprises the neutral Goldstone mode $G^0$, which becomes the longitudinal component of the Z boson, the Majoron $J$, and a physical CP-odd Higgs boson $A$. The resulting mass spectrum and phenomenology have been extensively studied in the literature \cite{FermiophobicHiggs, MAD1998htm}.

However, one might intuitively expect that the singular limit $\kappa \to 0$ removes the phase constraint, thereby restoring the possibility of SCPV. This intuition is incorrect. Analogous to the symmetry enhancement patterns in the 2HDM (see Table~5 of Ref.~\cite{Branco:2011iw}), in this limit the global symmetry is enhanced from $U(1)_L$ to two independent continuous symmetries:
\begin{equation}
U(1)_L \longrightarrow U(1)_{\sigma} \times U(1)_{\Delta},
\end{equation}

under which the fields transform as:
\begin{align}
U(1)_\sigma: \quad &\sigma \to e^{2i\alpha_\sigma}\sigma, \\
U(1)_\Delta: \quad &\Delta \to e^{-2i\alpha_\Delta}\Delta.
\end{align}
The vacuum phases transform as $\theta'_1 = \theta_1 + 2\alpha_\sigma$ and $\theta'_3 = \theta_3 - 2\alpha_\Delta$. By an appropriate rephasing, $\alpha_\sigma = -\theta_1/2$ and $\alpha_\Delta = \theta_3/2$ one can set $\theta_1=\theta_3=0$. Hence the VEV phases are unphysical and cannot signal spontaneous CP violation.

Although this limit does not yield physical CP violation, it drastically alters the mass spectrum. As shown in Appendix~\ref{app:massmatrix}, once the vacuum is rotated to be real, all quartic coupling contributions cancel out exactly, leaving every element of the CP-odd mass matrix $\mathcal{M}^2_I$ strictly proportional to $\kappa$. Consequently:

\begin{equation}
\lim_{\kappa \to 0} \mathcal{M}^2_I = \mathbf{0}_{3\times3}.
\end{equation}

The vanishing of the mass matrix implies the existence of additional massless modes. This is physically required by Goldstone's theorem: since the vacuum now spontaneously breaks two independent global symmetries ($U(1)_\sigma$ and $U(1)_\Delta$) rather than just one, a second physical Goldstone boson must necessarily appear in the spectrum.

This implies three massless modes in this limit: the neutral would-be Goldstone $G^0$, which provides the longitudinal component of the Z boson, the Majoron $J$, and an additional physical Goldstone mode. In the $\kappa \to 0$ limit the CP-odd scalar $A$ becomes massless and can be identified with this extra Goldstone; we denote it $\mathcal{G}$ (the “Gumooron”). The appearance of an extra massless degree of freedom renders this limit phenomenologically disfavored due to strict constraints on the Z-boson invisible decay width \cite{ParticleDataGroup:2024cfk}, justifying our focus on the generic coupled case where $\kappa \neq 0$.

Consequently, the scalar sector cannot source CP violation in any parameter space region. Therefore, beyond the standard CKM phase in the quark sector, any new CP violation required for leptogenesis must arise explicitly from complex Yukawa couplings in the neutrino sector \cite{Buchmuller:2005eh}. This eliminates spontaneous scalar VEV phases as free parameters, rendering the model more predictive.

\section{Discussion and Implications}
\label{sec:discussion}
\subsection{Comparison with the 2HDM}
It is instructive to compare this result with the Two-Higgs-Doublet Model (2HDM), where SCPV is a well-established possibility \cite{Lee:1973iz, Branco:2011iw}. The difference stems from the distinct symmetry structures of the vacuum:
\begin{itemize}
    \item \textbf{2HDM:} The model contains two representations with identical quantum numbers ($Y=1/2$). The relative phase between the two doublets cannot be removed if the potential softly breaks the basis symmetry, allowing for a physical complex phase in the vacuum. The explicit conditions for SCPV in the 2HDM, formulated in terms of physical masses and couplings, are discussed in detail in Sec.~3 of Ref.~\cite{Grzadkowski:2016scpv}.
    \item \textbf{123 Model:} The trilinear term $\kappa$ couples three distinct representations (singlet, doublet, triplet). The phase constraint $\theta_1 + \theta_3 = n\pi$ is rigid when $\kappa\neq 0$. The specific charge assignment of the $U(1)_L$ symmetry ($+2$ for singlet, $-2$ for triplet) allows the rotation parameter $\alpha$ to exactly compensate the phases of both fields simultaneously, rotating the complex vacuum to a real one. In the decoupling limit $\kappa = 0$, the enhanced symmetry structure $U(1)_\sigma \times U(1)_{\Delta}$ provides sufficient freedom to remove all phases independently.
\end{itemize}
This highlights a fundamental difference between models with identical representations (doublets) and models mixing distinct multiplets (singlet-doublet-triplet).

\subsection{Implications for Leptogenesis}

This constraint directly impacts leptogenesis scenarios in the 123 model \cite{Ma:1998, Hambye:2005tq, Branco:2012zb}. Since the standard CKM phase is insufficient to explain the baryon asymmetry \cite{GavelaCKM1, GavelaCKM2}, new CP violation sources are required. However, as the scalar vacuum is strictly CP-conserving, it cannot source this asymmetry. Consequently, the CP violation required for baryogenesis must arise explicitly from complex neutrino Yukawa couplings \cite{Buchmuller:2005eh}.

This result simplifies the phenomenological analysis by eliminating spontaneous scalar phases as free parameters. Relevant CP-violating phases are thus encoded entirely in the neutrino Yukawa sector.

\section{Conclusions}
\label{sec:conclusions}

We have analyzed the vacuum structure of the 123 model to determine whether spontaneous CP violation can arise in the scalar sector. Our results establish that the scalar vacuum is strictly CP-conserving across the entire parameter space. For $\kappa \neq 0$, the minimization conditions enforce a phase alignment $\theta_1 + \theta_3 = n\pi$, which combined with the global $U(1)_L$ symmetry allows all VEVs to be simultaneously rotated to real values. In the decoupling limit $\kappa \to 0$, the symmetry is enhanced to $U(1)_{\sigma} \times U(1)_{\Delta}$, rendering the vacuum phases unphysical gauge artifacts. This limit is further disfavored by the appearance of an additional massless Goldstone boson.

This result highlights a fundamental difference between triplet-extended models and doublet-based extensions such as the 2HDM, where SCPV is a well-established feature. Consequently, any CP violation required for leptogenesis in the 123 model must be introduced explicitly through complex Yukawa couplings in the neutrino sector, rather than emerging spontaneously from the scalar vacuum. This constraint simplifies the phenomenological analysis of the model by eliminating spontaneous VEV phases as free parameters.

\begin{acknowledgments}
This work is dedicated to Professor Marco Aurelio Díaz on the occasion of his retirement. We honor his distinguished career and his lasting contribution to the development of theoretical physics in Chile and to this work. Thanks for everything professor. We acknowledge support from ANID FONDECYT grant No. 11251121.
\end{acknowledgments}

\appendix

\section{Scalar Mass Structure and Zero Modes}
\label{app:massmatrix}

The neutral scalar mass matrix $\mathcal{M}^2$ is a $6\times 6$ symmetric matrix in the basis of fluctuations $(R_1, R_2, R_3, I_1, I_2, I_3)$.
\subsection{CP Conservation in the General Case}
\label{app:A1}

The potential mixing between CP-even ($R$) and CP-odd ($I$) sectors is governed by the off-diagonal block $\mathcal{M}^2_{RI}$. Explicit calculation of the mass matrix elements reveals that they are proportional to the imaginary parts of the vacuum expectation values, appearing as differences between the VEVs and their complex conjugates. 
To illustrate this dependence, representative mixing terms involving the trilinear coupling $\kappa$ are given by:

\begin{align}
(\mathcal{M}^2_{RI})_{12} &= -\frac{i}{2} \kappa v_2 (v_3 - v_3^*), \\
(\mathcal{M}^2_{RI})_{22} &= -\frac{i}{2} \kappa (v_1 v_3 - v_1^* v_3^*).
\end{align}

As demonstrated in Sec.~\ref{sec:absence_scpv}, the minimization conditions combined with the global $U(1)_L$ symmetry allow us to rotate the vacuum into a basis where all VEVs are real ($v_i = v_i^*$). Consequently, all terms of the form $(v_i - v_i^*)$ and $(v_i v_j - v_i^* v_j^*)$ vanish identically. Thus, the mass matrix becomes block-diagonal, $\mathcal{M}^2 = \text{diag}(\mathcal{M}^2_R, \mathcal{M}^2_I)$, confirming that the physical eigenstates have definite CP parity.

\subsection{The CP-Odd Sector and the Singular Limit}
In the singular limit $\kappa \to 0$, the enhanced global symmetry $U(1)_\sigma \times U(1)_\Delta$ allows us to rotate the vacuum into a basis where all VEVs are strictly real ($v_i = v_i^*$). This has two crucial consequences for the full $6\times 6$ neutral mass matrix. First, as shown in Appendix~\ref{app:A1}, real VEVs ensures that the off-diagonal mixing block vanishes identically ($\mathcal{M}^2_{RI} = 0$). Second, it dictates the structure of the remaining CP-odd block $\mathcal{M}^2_I$.

Crucially, the raw analytical expressions for $\mathcal{M}^2_I$ contain terms proportional to the quartic couplings ($\lambda_i, \beta_i$) multiplied by factors that vanish for real VEVs (e.g., $v_i - v_i^*$ or quadratic combinations summing to zero). Thus, the requirement of a real vacuum—enabled here by the symmetry enhancement—algebraically eliminates all contributions from the quartic sector. The only surviving terms are those proportional to the coupling $\kappa$.

The explicit surviving elements of $\mathcal{M}^2_I$ are:
\begin{align}
(\mathcal{M}^2_I)_{11} &= \frac{\kappa v_2^2 v_3}{2 v_1}, \\
(\mathcal{M}^2_I)_{22} &= 2 \kappa v_1 v_3, \\
(\mathcal{M}^2_I)_{33} &= \frac{\kappa v_1 v_2^2}{2 v_3}, \\
(\mathcal{M}^2_I)_{12} &= \kappa v_2 v_3, \\
(\mathcal{M}^2_I)_{13} &= \frac{1}{2} \kappa v_2^2, \\
(\mathcal{M}^2_I)_{23} &= \kappa v_1 v_2.
\end{align}

These exact expressions demonstrate that every non-zero element of the pseudoscalar mass matrix is strictly proportional to $\kappa$. Consequently, since $\mathcal{M}^2_{RI}$ is already zero due to the real vacuum, the vanishing of $\kappa$ implies that the CP-odd sector becomes entirely massless:

\begin{equation}
\lim_{\kappa \to 0} \mathcal{M}^2_I = \mathbf{0}_{3\times3}.
\end{equation}

This confirms the existence of three massless modes in the decoupling limit: the standard Goldstone $G^0$ (absorbed by the $Z$), the Majoron $J$, and the extra Gumooron $\mathcal{G}$ arising from the symmetry enhancement.

\bibliography{bibliography} 

\end{document}